\documentclass[italian,english]{revtex4}
\usepackage[T1]{fontenc}
\usepackage[latin1]{inputenc}
\usepackage{array}
\usepackage{amsmath}
\usepackage{graphicx}
\usepackage{amssymb}

\makeatletter

\providecommand{\tabularnewline}{\\}

\usepackage{babel}
\makeatother
\begin{document}

\title{Observational constraints on the linear fluctuation growth rate}

\author{Cinzia Di Porto}

\affiliation{Dipartimento di Fisica {}``E. Amaldi'', Università degli Studi
{}``Roma Tre''}

\author{Luca Amendola}

\affiliation{INAF/Osservatorio Astronomico di Roma, Via Frascati 33\\
 00040 Monte Porzio Catone (Roma), Italy}

\date{\today{}}

\begin{abstract}
Several experiments in the near future will test dark energy through
its effects on the linear growth of matter perturbations. It is therefore
important to find simple and at the same time general parametrizations
of the linear growth rate. We show that a simple fitting formula that
generalizes previous expressions reproduces the growth function in
models that allow for a growth faster than standard, as for instance
in scalar-tensor models. We use data from galaxy and Lyman-$\alpha$
power spectra to constrain the linear growth rate. We find $\gamma=0.6{}_{-0.3}^{+0.4}$
for the growth rate index and $\eta=0.0_{-0.2}^{+0.3}$ for the additional
growth parameter we introduce.
\end{abstract}
\maketitle

\section{Introduction}

After several years from the first works \cite{riess,perl}, the evidence
of dark energy (DE) still rests primarily on background quantities
like the luminosity distance and the angular diameter distance \cite{snls,wmap3,sdss_bao}.
Only recently the cross-correlation of the integrated Sachs-Wolfe
effect with the large scale structure yielded an independent proof
of the existence of dark energy that rely on the linear growth of
the gravitational potential \cite{isw}. In the future, galaxy and
Lyman-$\alpha$ power spectra at high redshift and weak lensing surveys
from ground and from space will offer the opportunity to test competing
dark energy models to a very high precision using a mix of background,
linear and non-linear indicators \cite{DUNE,JDEM}.

In order to use the growth rate of linear fluctuations as a test of
DE, it is however necessary to formulate simple expressions that embody
a large class of models, so as to provide observers with a practical
tool to analyse the data. This procedure has proved most convenient
with the equation of state of DE, whose parametrizations afford the
community to quickly compare the different experiments and to optimize
the design of further surveys (see eg \cite{park}).

Since many years it has been known that a good approximation to the
growth rate of the linear matter density contrast $\delta$ in standard
models of gravity and dark matter is given by the simple expression
\cite{lahav,wang}\begin{equation}
s\equiv\frac{\delta'}{\delta}\approx\Omega_{m}^{\gamma}\,,\label{eq:rate1}\end{equation}
where the prime stands for derivative with respect to $\log a$ and
where $\gamma\approx0.55$. In \cite{amque} it was found that this
growth rate also accounts for coupled dark energy at small $z$ and
small coupling $\beta$ (see definition below) with $\gamma\approx0.56(1-2.55\beta^{2})$.
The formula works quite well also in the case in which DE is described
by a mildly varying EOS $w(z)$ when generalized with \cite{linder}
$\gamma\approx0.55+0.05(1+w(z=1))$.

However, this formula has an obvious drawback: assuming a constant
or weakly varying $\gamma>0$, it implies $s\le1$ at all epochs since
in standard cosmology the matter density parameter is always $0\le\Omega_{m}\leq1$.
More exactly, if $\gamma$ is a constant, then $s$ is always either
smaller or larger than unity. Therefore, this fit is unable to test
for deviations from the standard paradigm of fluctuation growth, which
assumes $s=1$ for the matter-dominated universe at high redshifts.
This assumption is of course dangerous in view of the multitude of
dark energy or modified gravity models currently being studied and
should not be taken for granted without serious scrutiny.

In this paper we show first that the $s>1$ behavior takes place in
one of the simplest class of modified gravity theories, namely scalar-tensor
models or their Einstein frame counterparts, and then we introduce
a simple new fitting formula that generalizes eq. (\ref{eq:rate1}).
We note that a similar faster-than-standard growth has been found
also in TeVeS modified gravity models \cite{teves}. Finally, we compare
our fitting formula with the (scanty) data available at the present.

\section{The growth function in scalar-tensor theories}

In this section we show that generically $s>1$ at early times and
$s<1$ at late times in a simple class of modified gravity theories.

Let us consider a generic scalar-tensor model in Einstein frame. It
is well-known that in the Einstein frame, the frame in which the gravitational
equations are in the Einsteinian form while matter is not conserved,
\foreignlanguage{italian}{the scalar degree of freedom of a scalar-tensor
theory acts as a new force on matter, as expressed by the new conservation
equations (here we use subscripts $c$ for CDM, $b$ for baryons,
$\phi$ for the scalar field, and $\gamma$ for radiation ) \cite{DGG,wett95,coupled}\begin{eqnarray}
T_{\nu;\mu(c)}^{\mu} & = & -C_{c}(\phi)T_{(c)}\phi_{;\mu}\,,\label{eq:C2a}\\
T_{\nu;\mu(b)}^{\mu_{\ }} & = & -C_{b}(\phi)T_{(b)}\phi_{;\mu}\,,\label{eq:C2b}\\
T_{\nu;\mu(\phi)}^{\mu} & = & [C_{b}(\phi)T_{(b)}+C_{c}(\phi)T_{(c)}]\phi_{;\mu}\,,\label{eq:C2c}\\
T_{\nu;\mu(\gamma)}^{\mu} & = & 0\,,\label{eq:C2c1}\end{eqnarray}
where for generality we assumed that the coupling functions depend
on the species (while radiation remains uncoupled because it is conformally
invariant). In a FRW metric with scale factor $a$ these equations
become (we assume flat space throughout)}

\begin{eqnarray}
\ddot{\phi}+3H\dot{\phi}+V_{,\phi} & = & \kappa(\beta_{c}\rho_{c}+\beta_{b}\rho_{b})\,,\nonumber \\
\dot{\rho_{c}}+3H\rho_{c} & = & -\kappa\beta_{c}\rho_{c}\dot{\phi}\,,\nonumber \\
\dot{\rho_{b}}+3H\rho_{b} & = & -\kappa\beta_{b}\rho_{b}\dot{\phi}\,,\nonumber \\
\dot{\rho_{\gamma}}+4H\rho_{\gamma} & = & 0\,,\nonumber \\
3H^{2} & = & \kappa^{2}(\rho_{b}+\rho_{c}+\rho_{\phi}),\label{eq:C2sist1}\end{eqnarray}
where $\kappa^{2}=8\pi G$, $\beta_{c}=C_{c}/\kappa$, $\beta_{b}=C_{b}/\kappa$,
$H=\dot{a}/a$ (note that we use a coupling $\beta$ which is $\sqrt{2/3}$
the $\beta$ used in ref. \cite{coupled}). Then we have immediately
\begin{equation}
\rho_{c,b}=\rho(0)_{c,b}a^{-3}\exp\left\{ -\int\beta_{c,b}(\phi)d\phi\right\} \,.\label{eq:C2d}\end{equation}
To simplify the analysis and to satisfy local gravity constraints
we put from now on $\beta_{b}=0$ and $\beta_{c}=\beta=const$.

As it has been shown in refs. \cite{coupled}, for $\beta<\sqrt{3/2}$
the standard matter era that precedes the final acceleration is replaced
in this coupled model by an epoch in which the energy density fractions
$\Omega_{m},\Omega_{\phi}$ of matter and field are constant and equal
to \begin{equation}
\Omega_{\phi}=\frac{2}{3}\beta^{2}\,,\end{equation}
and $\Omega_{m}=1-\Omega_{\phi}$. During this epoch one has $\phi'=2\beta$
(the prime stands for $d/d\log a$) and the scale factor grows as
$a\sim t^{\frac{2}{3(1+w_{e})}}$ \foreignlanguage{english}{with}
$w_{e}=2\beta^{2}/3$ \foreignlanguage{english}{(these values are
approximated since are obtained neglecting both baryons and radiation).}
This new matter era has been denoted as $\phi$MDE in previous works.
This occurs when the potential is negligible with respect to the field
kinetic energy. Since the potential is dominating the final accelerated
epoch, it is clear that the $\phi$MDE generically will take place
before acceleration and, of course, after the radiation era. This
in fact is what has been observed in several numerical and analytical
investigations, for instance in the case of exponentials and inverse
power-law potentials $V(\phi)=A\phi^{-n}$ \cite{aq}. In the Jordan
frame, where matter is conserved, the $\phi$MDE corresponds to the
standard solution of the Brans-Dicke original theory (which is derived
in absence of a potential)\foreignlanguage{english}{\begin{equation}
a_{J}\sim t^{\frac{2+2\omega}{4+3\omega}}\,,\end{equation}
upon the substitution\begin{equation}
\beta^{2}=\frac{1}{2(3+2\omega)}.\end{equation}
Therefore, the} $\phi$\foreignlanguage{english}{MDE is quite a generic
feature of scalar-tensor models and it also shows up in some} $f(R)$
\foreignlanguage{english}{models \cite{apt}. We now show that during}
$\phi$\foreignlanguage{english}{MDE the growth of fluctuations is
faster that in a standard matter era. In ref. \cite{linnonlin} it
has been shown that the perturbation equations in the sub-horizon
regime is}\begin{equation}
\delta_{c}^{\prime\prime}+\left(1+\frac{\mathcal{H}^{\prime}}{\mathcal{H}}-\beta_{c}\phi'\right)\delta_{c}^{\prime}-\frac{3}{2}(\gamma_{cc}\delta_{c}\Omega_{c}+\gamma_{bc}\delta_{b}\Omega_{b})=0\,,\label{eq:delta''CDM}\end{equation}
where again the prime stands for derivation with respect to $\log a$,
and where $\gamma_{ij}=1+2\beta_{i}\beta_{j}$ and $\mathcal{H}$
is the conformal Hubble function $\mathcal{H}\equiv aH$. We assume
the baryon component to be negligible. Then the fluctuation equation
can be solved analytically during the $\phi$MDE:\begin{equation}
\delta\sim a^{1+2\beta^{2}}\,,\end{equation}
from which it appears that $s=1+2\beta^{2}$. In the Jordan frame,
the growth rate is instead\begin{equation}
\delta_{J}\sim a^{\frac{2+\omega}{1+\omega}}\sim a^{\frac{1+2\beta^{2}}{1-2\beta^{2}}}\,,\end{equation}
which also gives a rate larger than unity. When the $\phi$MDE ends
and acceleration takes over, the rate $s$ declines steadily to zero
as in standard cases. Therefore, as we anticipated, $s$ goes from
a value larger than unity to a value smaller than unity.

\section{A generalized fit}

We now proceed to find a convenient fit to the full evolution of $\delta(a)$
for the coupled models introduced in the previous section. In the
standard scenario $\delta$ obeys the equation\begin{equation}
\delta^{\prime\prime}(\alpha)+(1+\frac{\mathcal{H}^{\prime}}{\mathcal{H}})\delta^{\prime}(\alpha)-\frac{3}{2}\Omega_{m}\delta(\alpha)=0,\label{eq:g2}\end{equation}
where\begin{equation}
\frac{\mathcal{H}^{\prime}}{\mathcal{H}}=-\frac{1}{2}(1+3w_{\phi}(\alpha)\Omega_{\phi}(\alpha))=-\frac{1}{2}(1+\frac{\Omega'_{m}}{\Omega_{m}})\end{equation}
 The solution can then be approximated as\begin{equation}
\delta(\alpha)=e^{\int_{0}^{\alpha}d\alpha^{\prime}\Omega_{m}(\alpha^{\prime})^{\gamma}}\,.\label{eq:g,param}\end{equation}
In our modified gravity eq. (\ref{eq:g2}) becomes\begin{equation}
\delta^{\prime\prime}+\left(1+\frac{\mathcal{H}^{\prime}}{\mathcal{H}}-\beta\phi'\right)\delta^{\prime}-\frac{3}{2}\Omega_{m}(1+2\beta^{2})\delta=0\,.\label{eq:g''+beta}\end{equation}
One simple possibility would be to generalize (\ref{eq:g,param})
as\begin{equation}
\delta(\alpha)=e^{\int_{0}^{\alpha}d\alpha^{\prime}\Omega_{m}(\alpha)^{\gamma}(1+c\beta^{2})}\,,\label{eq:g,param2}\end{equation}
with $c$ a parameter to be determined by a least square fit. The
choice of a $\beta^{2}$ behavior is suggested by the fact that the
$\phi$MDE depends only on $\beta^{2}$. However, this new fit is
not very practical because it contains the function $\Omega_{m}(\alpha)$
that should be obtained by numerically integrating the background
equations and therefore depends on the field potential. To overcome
this difficulty, we propose to use instead the \emph{standard expression}
for $\Omega_{m}$ : \begin{equation}
\Omega_{m}^{(s)}(a)=\frac{\Omega_{m,0}}{\Omega_{m,0}+(1-\Omega_{m,0})a^{-3\hat{w}}}\,,\label{eq:om,std}\end{equation}
where $\hat{w}=(\log a)^{-1}\int w(a)da/a$ and the subscript $0$
denotes the present time. For the coupled dark energy model we are
considering here, we approximate $w(z)\approx w_{\phi}(z=0)$; although
one could easily expand $w(z)$ to higher orders, our approximation
is sufficient to show that our generalized fit works well. Therefore
we define the rate\begin{equation}
s\equiv\Omega_{m}^{(s)}(\alpha)^{\gamma}(1+c\beta^{2})\,.\label{eq:sfit}\end{equation}
In this way, the growth rate can be parametrized by $\Omega_{m,0},\gamma$,
and the combination $\eta\equiv c\beta$, plus the parameters that
enter $w(z)$. Now, even in the limit $\Omega_{m}\to1$ one has $s\not=1$.
In the next section we show that this generalized fit is indeed a
good approximation. Since we know that during the $\phi$MDE (i.e.
at high $z$, for which $\Omega_{m}^{(s)}\approx1$) one has $s=1+2\beta^{2}$
we can anticipate that the result will be close to $c\approx2$.

Concluding this section we note that eq. (\ref{eq:sfit}) should be
seen for what it is, i.e. a phenomenological fit. The relation of
$\Omega_{m,0}$, $w(z)$, $\gamma$ and $\eta$ to the underlying
theory will of course depend on the theory itself. For instance, the
identification of $\Omega_{m,0}$ with the presently clustered mass
in galaxies and clusters of galaxies is actually a model-dependent
assumption; if gravity is not standard this assumption is likely to
be incorrect. All we are assuming here is that the Friedmann equation
can be written as the sum of two components, one that dilutes as $\Omega_{m,0}a^{-3}$
and the other as $(1-\Omega_{m,0})a^{-3(1+\hat{w})}$; if the gravitational
equations are not standard, one has to define $\Omega_{m}(a)$ such
that the above parametrization is still valid. The advantage of using
(\ref{eq:om,std}-\ref{eq:sfit}) is that both background and linear
growth are fitted by the same expression for $w(a)$; that is, once
one adopts a prescription for $w(a)$ one can fit all the data by
simply adding the two parameters $\gamma$ and $\eta$ (plus possibly
further parameters to account for the anisotropic stress, see eg.
\cite{aks}). Of course, in principle one could proceed in different
ways: for instance, one could parametrize $\Omega_{m}(a)$ so that
values larger than unity in the past were allowed so as to force $s>1$.
Trivially, in fact, our parametrization above could be written equivalently
defining a new density $\Omega_{m}\equiv\Omega_{m}^{(s)}(1+\eta)^{1/\gamma}$;
this density parameter would however not be the same that appears
in the Friedmann equation.

\section{Comparing the fit to the numerical results.}

We solved numerically the background equations of the system (\ref{eq:C2sist1})
choosing an exponential form for the potential $V(\phi)$ and neglecting
the fraction of baryons and radiation. Then we solved numerically
the differential equation (\ref{eq:g''+beta}), thus obtaining a solution
(that we denote $\delta_{exact}$) which depends on the value of the
coupling constant $\beta$. For $\beta$ ranging between 0 and 1/2,
we found that the values of the parameters $\gamma$, $c$ appearing
in (\ref{eq:sfit}), which produce the best fit to $\delta_{exact}$
are $\gamma=0.56$ and $c=2.1$. Our best fit is therefore \begin{equation}
\delta_{fit}(\alpha,\beta)\equiv e^{\int_{0}^{\alpha}d\alpha^{\prime}\Omega_{m}^{(s)}(\alpha)^{0.56}(1+2.1\beta^{2})}\,,\label{eq:genfitbeta}\end{equation}
where we remark again that we use the standard expression for $\Omega_{m}(a)$.

\begin{figure}[t]

\begin{centering}\includegraphics[width=15cm,height=6cm,keepaspectratio]{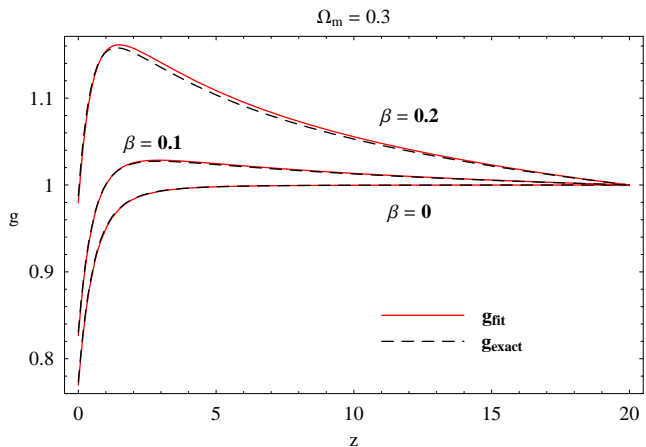}\par\end{centering}

\caption{\label{fig:compare_fit_gexact}We compare the functions $g_{fit}\equiv\delta_{fit}/a$
(red solid curves), given by the fitting formula (\ref{eq:g,param2})
with the best fit (parameters $\gamma$=0.56, $c=2.1$) for two different
values for $\beta$, to the exact solutions $g_{exact}\equiv\delta_{exact}/a$
(black dashed curves) of the differential equation (\ref{eq:g2})
for the growth rate. The curve for $\beta=0$ also gives the standard
best fit (i.e. for $c=0$). All curves are normalized at unity at
$z=20$. The cases $\beta=0,0.2$ are for an exponential potential,
the case $\beta=0.1$ for $V\sim1/\phi$.}

\end{figure}

This new function is really a good approximation to the exact solution
$\delta_{exact}$, for different values of the coupling constant $\beta$,
as one can see in Fig.\ref{fig:compare_fit_gexact} where the curves
of the growth factor $g\equiv\delta/a$ for two different $\beta$
and the corresponding $g_{fit}$ are plotted. In Fig.2 we present
the level of accuracy of the fitting formula. As it can be seen we
find fits to better than $\approx$1\% for different values of $\beta$.
Moreover, we find that the best fit values of the parameters do not
depend on the actual value of the present matter density $\Omega_{m,0}$.
We experimented also with an inverse power-law potential and found
that also in this case eq. (\ref{eq:genfitbeta}) is a good fit (see
curve for $\beta=0.1$ in Fig. \ref{fig:compare_fit_gexact}). Without
the $\eta$-correction the relative error $(\delta_{fit}-\delta_{exact})/\delta_{exact}$
becomes larger than 15\% already for $\beta=0.2$.

\begin{figure}[t]

\begin{centering}\includegraphics[width=15cm,height=7cm,keepaspectratio]{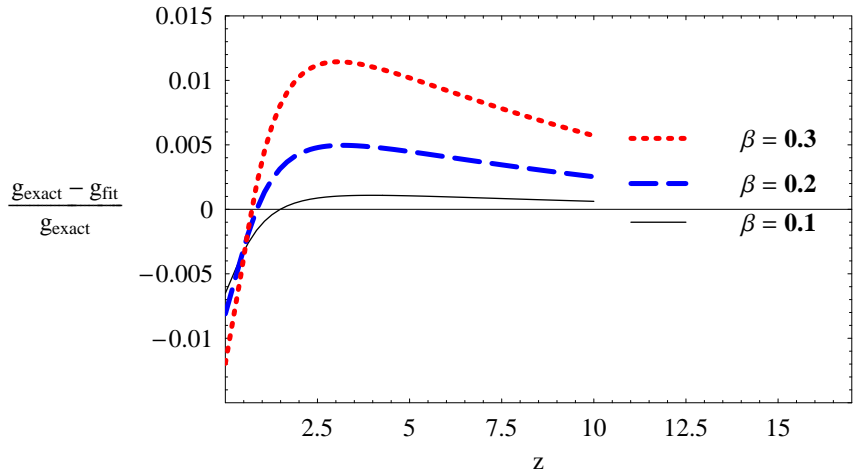}\par\end{centering}

\caption{Level of accuracy of the best fit to the exact solution for the growth
rate. For $\beta$ ranging between $0$ and $0.3$ the fits are better
than 1.2\%. Without the correction the relative errors would be larger
than 15\% already for $\beta=0.2$.}

\end{figure}

\section{Comparing the fit to the observations}

In the previous section we have seen that the expression \begin{equation}
s\equiv\Omega_{m}^{(s)}(\alpha)^{\gamma}(1+\eta)\,,\label{eq:genfit}\end{equation}
 where $\Omega_{m}^{(s)}(\alpha)$ is given by eq. (\ref{eq:om,std})
gives a good fit to the evolution of $\delta$ during both the decelerated
and accelerated regimes for coupled dark energy models with constant
$\beta<1/2$ if $\gamma\approx0.56$ and $\eta=2.1\beta^{2}$. Here
we take some preliminary steps towards comparing the fit (\ref{eq:genfit})
to the observations. An indication for a positive $\eta$ could signal
an attractive force additional to standard gravity as in a scalar-tensor
model; on the other hand one can speculate that a negative $\eta$
could be related to the slowed growth induced by a hot matter component. 

We consider the following data: $a$) Lyman-$\alpha$ power spectra
at an average redshift $z=2.125$, $z=2.72$ \cite{viel04}, $z=3$
\cite{macdon}; $b$) the normalization $\sigma_{8}$ inferred from
Lyman-$\alpha$ at $z$ ranging between 2 and 3.8 \cite{viel06};
$c$) galaxy power spectra at low $z$ from SDSS \cite{sdss} and
2dF \cite{2df}. From the three Lyman-$\alpha$ and the SDSS spectra
we estimate the ratios\begin{equation}
r(k_{i};z_{1},z_{2})=\frac{P(k_{i},z_{1})}{P(k_{i},z_{2})}\,,\end{equation}
for the values of $k_{i}$ for which there are tabulated value of
the spectra (or for interpolated values and errors when the tabulated
wavenumbers differ). For the $\sigma_{8}$ data we estimate the ratios
between successive values of $z$, \begin{equation}
r(z_{1},z_{2})=\frac{\sigma_{8}^{2}(z_{1})}{\sigma_{8}^{2}(z_{2})}\end{equation}
(note that ref. \cite{viel06} reports the values of $\sigma_{8}$
extrapolated at the present epoch). 

For the Lyman spectrum at $z=3$ and for 2dF ($z=0.15)$, the authors
of \cite{macdon,2df} give directly their estimation of the growth
rate, $s_{obs}=$$0.49$$\pm0.10$ for 2dF and $s_{obs}=$1.46$\pm0.29$
for the Lyman-$\alpha$ data. Then we compare the observations to
our fit by using the likelihood function\begin{equation}
L=N\exp\sum_{i}\left(-\frac{(r_{i,obs}-r_{i,teor})^{2}}{2\sigma_{i}^{2}}\right)\exp\sum_{j}\left(-\frac{(s_{j,obs}-s_{j,teor})^{2}}{2\sigma_{j}^{2}}\right)\,,\label{eq:L1}\end{equation}
where the errors $\sigma_{i}$ are obtained from the quoted errors
on $P(k)$ and $\sigma_{8}$ by standard error propagation.

As we will see the data available at the present are not sufficient
to set stringent limits to the growth function. Moreover, there are
several sources of possible systematic effects that we cannot account
for. For instance, the matter spectra derived from Lyman-$\alpha$
clouds are obtained through calibration (ie. bias correction) with
 $N$-body simulations; these simulations have been generated only
for a limited set of cosmological models. It is difficult to quantify
the impact of this limitation upon our results; the fact that we consider
\emph{ratios} of spectra from similar sources (eg Lyman-$\alpha$
clouds) might however alleviate the problem since one can expect that
the calibration errors are only weakly dependent on redshift. For
this reason we consider separately the ratios of the high-$z$ Lyman-$\alpha$
spectra to the low$-z$ SDSS galaxy spectra; our final results do
not take these into account.

\begin{figure}[t]

\begin{centering}\includegraphics[width=15cm,height=7cm,keepaspectratio]{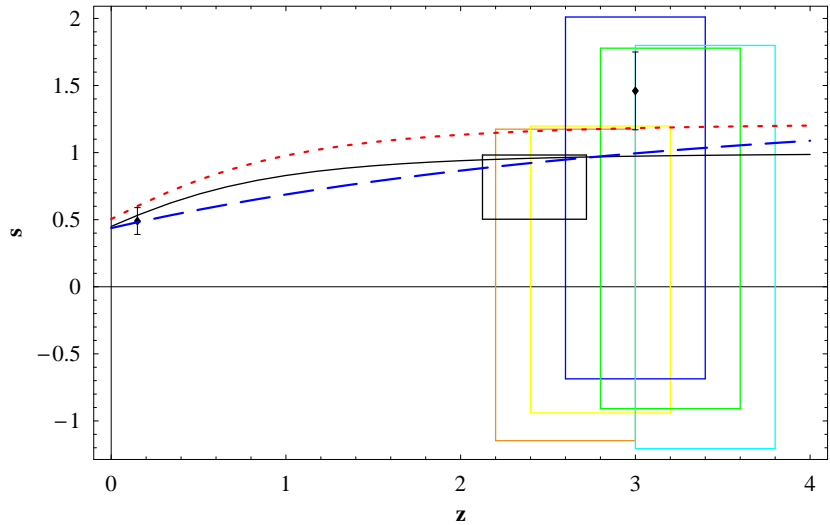}\par\end{centering}

\caption{\label{fig:data} Summary of experimental data for the growth rate
$s$, as detailed in Table I. The big coloured errorboxes represents
the ratios $\sigma(z_{1})/\sigma(z_{2})$ for various $z$ intervals
\foreignlanguage{english}{of ref. \cite{viel06}} (three additional
very large errorboxes have been excluded from the plot but not from
the analysis); the smaller black box represents the average spectral
ratio for the Lyman-$\alpha$ data of ref. \cite{viel04}. The two
points with errorbars are from \foreignlanguage{english}{ref. \cite{2df}}
\foreignlanguage{english}{and ref. \cite{macdon}}. The black solid
line is the $\Lambda$CDM model, the red curve is the coupled dark
energy model with $\Omega=0.2$ and $\beta=0.4$ (i.e. $\eta=0.34$)
and the dashed blue curve is the overall best fit $(\Omega_{m,0},w_{0},\gamma,\eta)=(0.05,-0.6,0.4,0.45)$.}

\end{figure}

The current observational situation is summarized in Fig. (\ref{fig:data})
(and the associated Table I), in which we plot the data we used in
this work, along with the $\Lambda$CDM growth rate and with our best
fit (see below). This figure gives a clear idea of the potential for
improvement in the observational estimation of the growth rate. 

We assume that the function $s$ depends on four parameters, $(\Omega_{m,0},w_{0},\gamma,\eta)$.
We assume also a flat prior $\Omega_{m,0}\in(0.05,0.4)$ and $w_{0}\in(-1,-0.6)$
which generously accounts for the supernovae constraints (neglecting
the phantom region). Our main result is contained in Fig. (\ref{contours}),
which displays the likelihood contour plots at 68\%,95 and 99.7\%
in the plane $(\gamma,\eta)$, marginalizing over $\Omega_{m,0},w_{0}$.
Remarkably, the best fit values practically coincide with the $\Lambda$CDM
prediction, $(\eta,\gamma)=(0,0.6)$. However the likelihood extends
considerably on both negative and positive $\eta$ and even negative
values of $\gamma$ are not excluded beyond 3$\sigma$. In Figs. (\ref{likgamma}-\ref{liketa})
we plot the marginalized 1D likelihoods for $\gamma$ and $\eta$.
The results are tabulated in Table II. The best fit values and $1\sigma$
errors are\begin{equation}
\gamma=0.60_{-0.30}^{+0.41}\,,\quad\eta=0.00_{-0.18}^{+0.28}\,.\end{equation}
 As we anticipated, the current data impose only very weak constraints
on $\gamma,\eta$ . For completeness, we also quote in Table II the
best fit and errors on $\gamma_{standard}$, i.e. assuming a standard
model in which $\eta=0$. Even in this case the likelihood distribution
for $\gamma$ remains very broad, although now negative values are
rejected at more than 3$\sigma$. Including the ratio of Lyman-$\alpha$
to SDSS power spectra has a minor effect on $\gamma$ and moves the
best fit of $\eta$ to $-0.2$.

Assuming $\eta<0.58$ at 2$\sigma$ we can derive an upper limit to
the coupling $\beta$ introduced in Sect. 2, \begin{equation}
\beta<0.52\end{equation}
(at 95\% c.l.). This limit is very weak when  compared to the CMB
limits \cite{aq} but it is nevertheless interesting since it is independent
and derived uniquely from the  growth rate at small redshifts.

\begin{table}

\begin{tabular}{|c|c|}
\hline 
$z$&

$s$
\tabularnewline
\hline
\hline 
\multicolumn{2}{|c|}{
ref. \cite{viel04}
}\tabularnewline
\hline 
2.125; 2.72&
0.50; 0.98\tabularnewline
\hline
\hline 
\multicolumn{2}{|c|}{ref. \cite{viel06}}\tabularnewline
\hline 
2.2; 3&
-1.147; 1.175\tabularnewline
\hline 
2.4; 3.2&
-0.94; 1.198\tabularnewline
\hline 
2.6; 3.4&
-0.686; 2.010\tabularnewline
\hline 
2.8; 3.6&

-0.908; 1.778
\tabularnewline
\hline 
3; 3.8&
-1.207; 1.799\tabularnewline
\hline
\hline 
\multicolumn{2}{|c|}{ref. \cite{macdon}}\tabularnewline
\hline
3&
1.46$\pm$0.49\tabularnewline
\hline
\hline 
\multicolumn{2}{|c|}{ref. \cite{2df}}\tabularnewline
\hline 
0.15&
0.49$\pm0.10$\tabularnewline
\hline
\end{tabular}

\caption{Summary of observational data. We report in the $z$ and $s$ columns
either the corresponding ranges or the central value and errors. For
the $\sigma_{8}$ data or ref. \cite{viel06} we chose to report the
errorboxes on $s$ obtained using the ratios at the given redshifts.}
\end{table}

\begin{table}

\begin{tabular}{|c|c|c|c|}
\hline 
&
$1\sigma$&
$2\sigma$&
$3\sigma$\tabularnewline
\hline
\hline 
$\eta$&

\begin{tabular}{c}
\tabularnewline
\tabularnewline
\end{tabular}$0.00{}_{-0.18}^{+0.28}$
&

$_{-0.38}^{+0.58}$
&

$_{-0.58}^{+1.1}$
\tabularnewline
\hline 
$\gamma$&

\begin{tabular}{c}
\tabularnewline
\tabularnewline
\end{tabular}\foreignlanguage{english}{$0.60{}_{-0.30}^{+0.41}$}
&

$_{-0.49}^{+0.97}$
&
$_{-0.74}^{+1.6}$\tabularnewline
\hline 
$\gamma_{standard}$&

\begin{tabular}{c}
\tabularnewline
\tabularnewline
\end{tabular}$0.60{}_{-0.26}^{+0.34}$
&

$_{-0.40}^{+0.77}$
&

${}_{-0.50}^{+1.4}$
\tabularnewline
\hline
\end{tabular}

\caption{Best fit and errors (marginalized over all other parameters). }
\end{table}

\begin{figure}[t]

\begin{centering}\includegraphics[width=15cm,height=7cm,keepaspectratio]{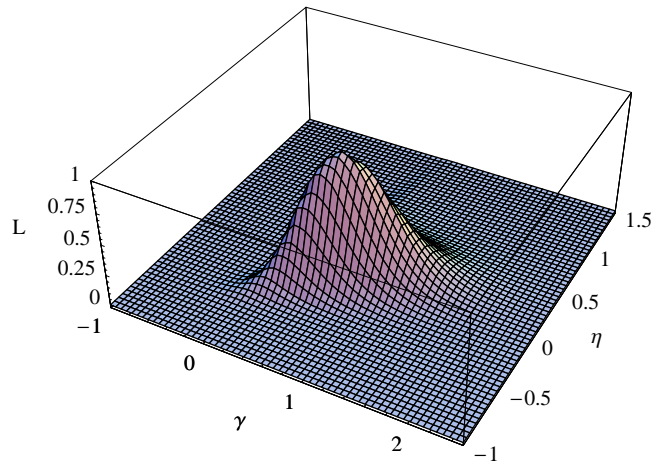}\par\end{centering}

\caption{Tridimensional likelihood function marginalized on $\Omega_{m,0}$
\foreignlanguage{english}{and} $w_{0}$ . The peak corresponds to
($\gamma,\eta$)=(0.6,0).}

\end{figure}
\begin{figure}[t]

\begin{centering}\includegraphics[width=15cm,height=7cm,keepaspectratio]{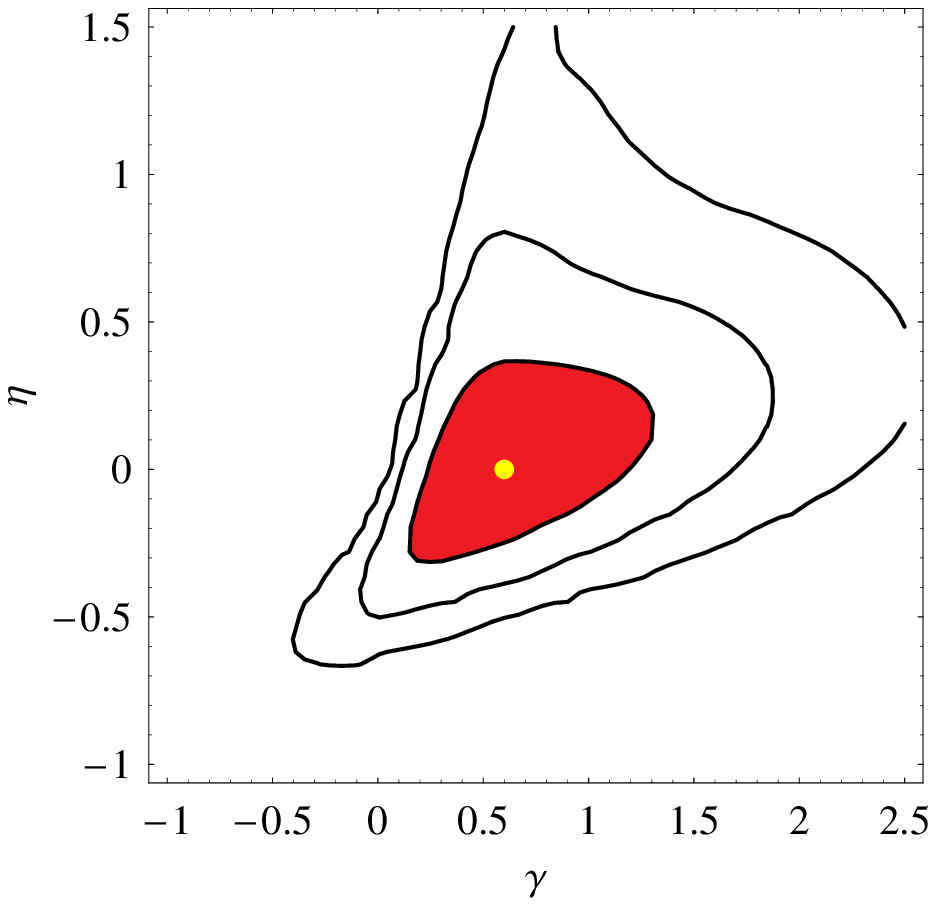}\par\end{centering}

\caption{\label{contours} Contour plot of the likelihood marginalized over
$\Omega_{m,0}$ and $w_{0}$. The contours, from inside to outside,
are at the 68\% (red zone), 95\%, 99.7\% confidence level. The dot
marks the peak $(\gamma,\eta)=(0.6,0)$.}

\end{figure}

\section{Conclusions}

The search for useful parametrizations of the dark dynamics is important
since as it has been shown several times every parametrization introduces
some arbitrariness in the way data are analysed \cite{bck}. In particular,
with the advent of models of dark energy based on modification of
Einstein's gravity, we have become aware of many possible trends,
both at the background and at the perturbation level, that are not
easily accounted for with earlier parametrizations. In this paper
we introduced a generalized form of parametrization of the growth
rate that allows for a rate $s\not=1$, i.e. faster or slower than
the standard matter-dominated growth. We show that this parametrization
is suitable to model the fluctuation growth in coupled dark energy
models and in scalar-tensor models. 

We have analysed the current data in search of observational constraints
on $\gamma,\eta$. Considering data from Lyman-$\alpha$ and galaxy
power spectra at various redshifts we have obtained (rather weak)
constraints on both parameters. The best fit turns out to be very
close to the $\Lambda$CDM predictions. Many future experiments based
on weak lensing or baryon oscillations will be able to estimate the
growth rate and other fluctuation parameters with much higher precision
\cite{HuLi,aks,ccm}. We expect therefore that the constraints derived
in this paper will soon be superseded by much more precise ones and
that new estimates of the growth factor will help in clarifying the
nature of dark energy.

\begin{acknowledgments}
We thank Enzo Branchini and Matteo Viel for useful discussions on
the data analysis.
\end{acknowledgments}
\begin{figure}[t]

\begin{centering}\includegraphics[width=15cm,height=7cm,keepaspectratio]{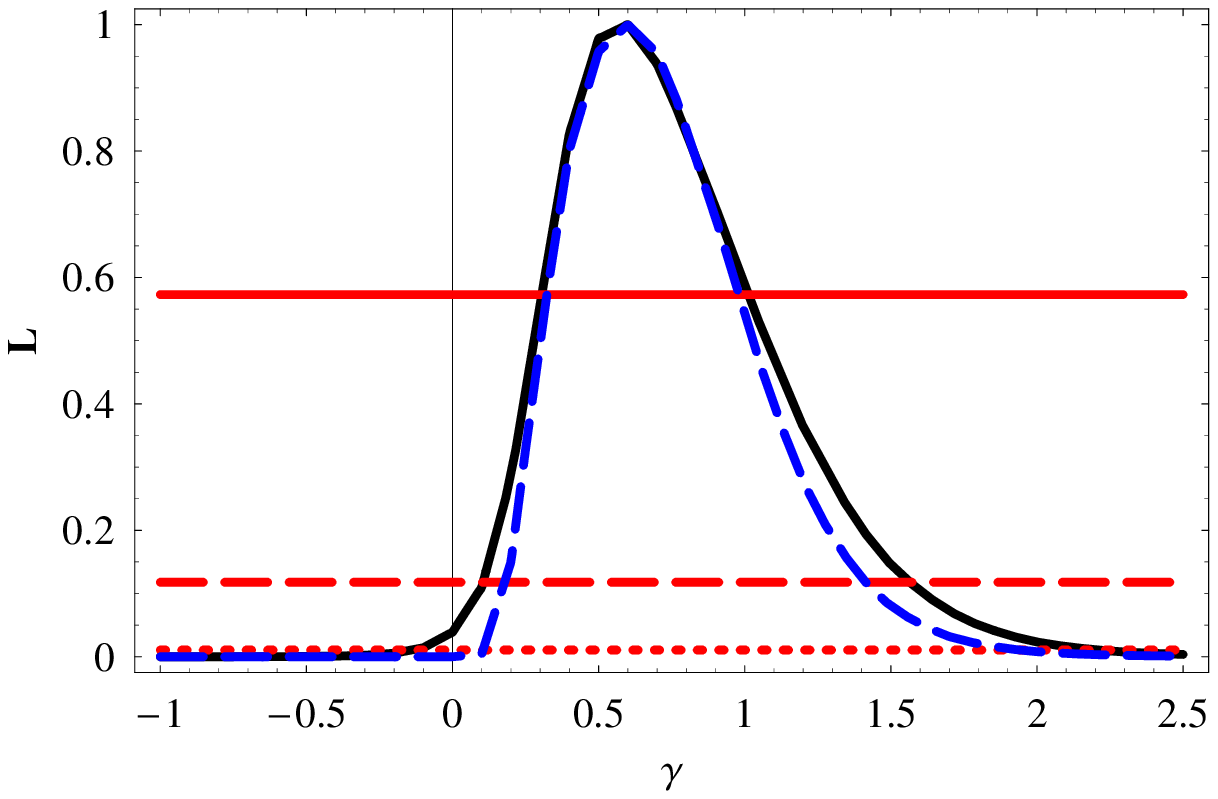}\par\end{centering}

\caption{\label{likgamma}Marginalized likelihood for $\gamma$ (solid line)
and for $\gamma_{standard}$ , i.e. fixing $\eta=0$ (dashed line).
The three horizontal lines represent the 68,95,99.7\% c.l. from top
to bottom.}

\end{figure}

\begin{figure}[t]

\begin{centering}\includegraphics[width=15cm,height=7cm,keepaspectratio]{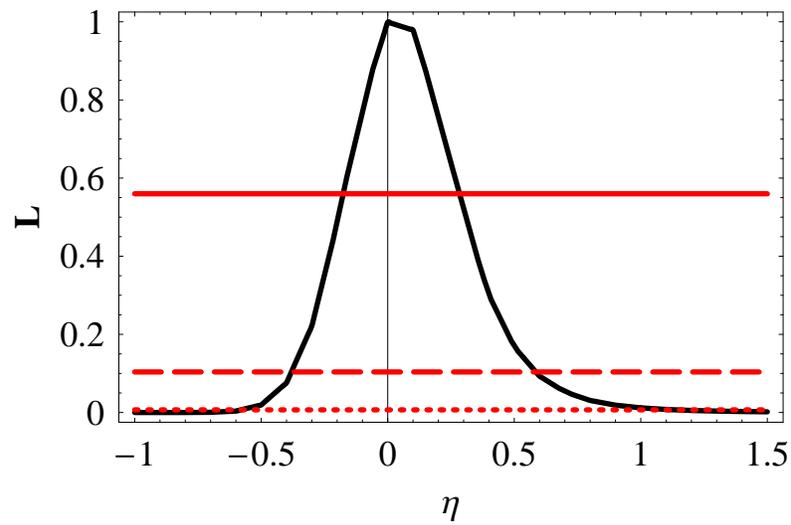}\par\end{centering}

\caption{\label{liketa} Marginalized likelihood for $\eta$. The three horizontal
lines represent the 68,95,99.7\% c.l. from top to bottom.}

\end{figure}

\end{document}